\documentclass[pra, twocolumn, final, reprint, amsmath, amssymb, superscriptaddress, showpacs, a4paper, aps, 10pt, hidelinks, numbers, sort&compress, floats, balancelastpage]{revtex4-1}

\usepackage{graphicx, color} 

\usepackage[utf8]{inputenc}

\newcommand{\1}{\ensuremath{\left|1 \right\rangle}}

\definecolor{britishracinggreen}{rgb}{0.0, 0.26, 0.15}
\definecolor{bulgarianrose}{rgb}{0.28, 0.02, 0.03}
\definecolor{darkred}{rgb}{0.90,0,0}
\definecolor{darkgreen}{rgb}{0,0.60,.2}
\definecolor{darkblue}{rgb}{0,0,1}
\definecolor{orange}{cmyk}{0,0.6,0.8,0}
\definecolor{lightblue}{rgb}{0.3,0.5,1}
\definecolor{lightgreen}{rgb}{0.4,0.80,.4}

\newcommand{\Li}{$^{6}$Li}

\newcommand{\beq}{\begin{equation}}
\newcommand{\eeq}{\end{equation}}
\newcommand{\bei}{\begin{itemize}}
\newcommand{\eei}{\end{itemize}}
\newcommand{\ben}{\begin{enumerate}}
\newcommand{\een}{\end{enumerate}}

\begin{document}

\title{Realization of a Cold Mixture of Fermionic Chromium and Lithium Atoms}



\author{E. Neri} 
\affiliation{Istituto Nazionale di Ottica del Consiglio Nazionale delle Ricerche (INO-CNR), 50019 Sesto Fiorentino, Italy}
\affiliation{European Laboratory for Nonlinear Spectroscopy (LENS), 50019 Sesto Fiorentino, Italy}
\affiliation{Dipartimento di Fisica e Astronomia, Università di Firenze, 50019 Sesto Fiorentino, Italy}

\author{A. Ciamei} 
\affiliation{Istituto Nazionale di Ottica del Consiglio Nazionale delle Ricerche (INO-CNR), 50019 Sesto Fiorentino, Italy}
\affiliation{European Laboratory for Nonlinear Spectroscopy (LENS), 50019 Sesto Fiorentino, Italy}
\affiliation{Dipartimento di Fisica e Astronomia, Università di Firenze, 50019 Sesto Fiorentino, Italy}

\author{C. Simonelli} 
\affiliation{Istituto Nazionale di Ottica del Consiglio Nazionale delle Ricerche (INO-CNR), 50019 Sesto Fiorentino, Italy}
\affiliation{European Laboratory for Nonlinear Spectroscopy (LENS), 50019 Sesto Fiorentino, Italy}
\affiliation{Dipartimento di Fisica e Astronomia, Università di Firenze, 50019 Sesto Fiorentino, Italy}

\author{I. Goti} 
\affiliation{Dipartimento di Fisica e Astronomia, Università di Firenze, 50019 Sesto Fiorentino, Italy}

\author{M. Inguscio} 
\affiliation{Istituto Nazionale di Ottica del Consiglio Nazionale delle Ricerche (INO-CNR), 50019 Sesto Fiorentino, Italy}
\affiliation{European Laboratory for Nonlinear Spectroscopy (LENS), 50019 Sesto Fiorentino, Italy}
\affiliation{Department of Engineering, Campus Bio-Medico University of Rome, 00128 Rome, Italy}

\author{A. Trenkwalder} 
\email[Corresponding author. E-mail: ]{andreas.trenkwalder@ino.cnr.it}
\affiliation{Istituto Nazionale di Ottica del Consiglio Nazionale delle Ricerche (INO-CNR), 50019 Sesto Fiorentino, Italy}
\affiliation{European Laboratory for Nonlinear Spectroscopy (LENS), 50019 Sesto Fiorentino, Italy}
\affiliation{Dipartimento di Fisica e Astronomia, Università di Firenze, 50019 Sesto Fiorentino, Italy}

\author{M. Zaccanti} 
\affiliation{Istituto Nazionale di Ottica del Consiglio Nazionale delle Ricerche (INO-CNR), 50019 Sesto Fiorentino, Italy}
\affiliation{European Laboratory for Nonlinear Spectroscopy (LENS), 50019 Sesto Fiorentino, Italy}
\affiliation{Dipartimento di Fisica e Astronomia, Università di Firenze, 50019 Sesto Fiorentino, Italy}


\begin{abstract}
We report on the production of a novel cold mixture of fermionic $^{53}$Cr and $^{6}$Li atoms delivered by two Zeeman-slowed atomic beams and collected within a magneto-optical trap (MOT). 
For lithium, we obtain clouds of up to $4 \,10^8$ atoms at temperatures of about $500\,\mu$K. A gray optical molasses stage allows us to decrease the gas temperature down to $45(5)\,\mu$K.
For chromium, we obtain MOTs comprising up to $1.5\, 10^6$ atoms. The availability of magnetically trappable metastable $D$-states, from which $P$-state atoms can radiatively decay onto, enables to accumulate into the MOT quadrupole samples of up to $10^7$ $^{53}$Cr atoms. After repumping  $D$-state atoms back into the cooling cycle, a final cooling stage decreases the chromium temperature down to $145(5)\,\mu$K.
While the presence of a lithium MOT decreases the lifetime of magnetically trapped $^{53}$Cr atoms, we obtain, within a 5 seconds duty cycle, samples of about  $4\, 10^6$ chromium and $1.5\,10^8$ lithium atoms. 
Our work provides a crucial step towards the production of degenerate Cr-Li Fermi mixtures. 
\end{abstract}

\maketitle

\section{INTRODUCTION}
\label{intro}
Ultracold Fermi gases represent nowadays a prominent platform for the implementation of model Hamiltonians and the exploration of a variety of many-body regimes, primarily in the context of superfluid pairing and magnetic ordering.  Notable examples are the study of the BCS-BEC crossover in fermionic mixtures with balanced or imbalanced spin populations \cite{Chevy2010, Radzihovsky2010, Zwerger2012}, and the disclosure of ferromagnetic \cite{Jo2009,Sanner2012,Valtolina2017,Amico2018,Scazza2019_ii} and anti-ferromagnetic \cite{Hart2015,Parsons2016, Boll2016,Cheuk2016} correlations within itinerant and localized repulsive fermion systems, respectively.
The introduction of a “heavy-light” mass asymmetry among the constituents of a fermionic mixture is considered as an appealing extension of this research field, that opens qualitatively new scenarios.
At the few-particle level, two-fermion mixtures are predicted to exhibit a rich variety of N-body bound states and scattering resonances, both for free-space and trapped systems. Among others, a proper mass imbalance leads to the existence of the three-, four- and five-body Efimov effect \cite{Efimov1971, Efimov1973, Castin2010,Bazak2017}, non-Efimovian trimer states, both in free space \cite{Kartavtsev2007, Endo2011, Bazak2017} and in reduced or mixed dimensions \cite{Nishida2008, Levinsen2009}, as well as few-body clusters in trapped configurations, see e.g. Ref.\,\cite{Blume2012} and references therein. 
At the many-body level, the natural mismatch of the Fermi surfaces of a two-component mixture is predicted to favor superfluidity beyond the standard Cooper pairing mechanism \cite{Fulde1964, Larkin1964, Sarma1963, Gubankova2003, Liu2003, Forbes2005, Iskin2006, Parish2007, Baranov2008,Baarsma2010,Gezerlis2009,Astrakharchik2012}  at experimentally attainable temperatures. Moreover, a number of exotic normal many-body phases, such as ordered crystalline states \cite{Petrov2007} and trimer Fermi gases \cite{Nishida2008b, Levinsen2009}, have been theoretically investigated.
Parallel to this, mass-imbalanced mixtures have recently emerged also as a promising framework for the quantum simulation of magnetic phenomena, arising in both itinerant \cite{Keyserlingk2011, Cui2013,Massignan2014} and localized \cite{Sotnikov2012,Sotnikov2013} fermion systems with short-range repulsive interactions.

In spite of the great interest for ultracold Fermi gases of chemically different species, their experimental investigation is still at a relatively early stage, only a very few degenerate Fermi systems with sizable mass imbalance being currently available: the $^{173}$Yb-$^6$Li \cite{Hara2011, Gupta2019}, the $^{40}$K-$^6$Li \cite{Taglieber2008,Wille2008} and the $^{161}$Dy-$^{40}$K \cite{Ravensbergen2018,Ravensbergen2018} mixtures. 
The first combination is primarily appealing for the creation of ground state polar molecules with both magnetic and electric dipole moment \cite{Micheli2006}. However, the extremely narrow nature of the predicted Feshbach resonances \cite{Brue2012}, combined with a small background scattering length \cite{Hara2011} makes the reach of the strongly interacting regime in such a system extremely challenging. 
Furthermore, the very large mass ratio $M_{Yb}/m_{Li} \sim 29$ would support, at strong coupling, the existence of Efimov cluster states \cite{Efimov1971, Efimov1973, Kartavtsev2007, Endo2011},  whose intrinsically lossy nature could strongly limit the stability of a resonant Yb-Li mixture. 

In this respect, the $^{40}$K-$^6$Li system appears as a much more promising platform. Thanks to the availability of interspecies Feshbach resonances \cite{Wille2008,Naik2011}, interesting results have indeed been obtained both in the many-body \cite{Trenkwalder2011,Kohstall2012,Cetina2015,Cetina2016} and in the few-body \cite{Jag2014,Jag2016} context. In particular, the large but not extreme mass ratio of such a mixture, $M_{K}/m_{Li} \sim 6.6$, has enabled to experimentally unveil a resonant enhancement of $p$-wave K-KLi atom-dimer interactions on the BEC side of a K-Li Feshbach resonance \cite{Jag2014}. This exotic few-body feature starkly differs from the phenomenology typical of the Efimov scenario, as it exhibits a purely elastic, universal character: as such, it may represent an unforeseen opportunity to investigate in the future many-body regimes of ultracold fermionic matter in presence of non-perturbative few-body correlations. Furthermore, within a confined geometry \cite{Levinsen2011}, this could enable to realize collisionally stable trimer Fermi gases with exotic topological properties \cite{Tajima2019}. 
However, the narrow character of the K-Li resonances, and especially the presence of non-zero two-body inelastic decay \cite{Naik2011}, render the investigation of ground-state properties with such a mixture extremely challenging. 

These complications will be hopefully overcome in the near future by the recently realized $^{161}$Dy-$^{40}$K \cite{Ravensbergen2018} mixture, on which wide resonances in the system ground state, immune to two-body losses, have been discovered, and strongly interacting conditions have been attained \cite{Ravensbergen2019}. Yet, the very dense spectrum of both intra- and inter-species resonances in Dy-Dy and K-Dy collisions, connected with the anisotropic nature of the interatomic interactions \cite{Petrov2012}, could complicate the production of highly degenerate samples and the controlled tuning of Dy-K elastic interactions. Further, the mass ratio $M_{Dy}/m_{K} \sim 4.0$ of such a mixture, significantly smaller than the K-Li one, may prevent, at least in three dimensions, the emergence of non-trivial few-body physics. 

Here, we report on the first important step towards the production of a novel Fermi mixture of $^{53}$Cr and $^6$Li atoms, namely the realization of magneto-optically trapped Cr-Li clouds in the cold regime. 
Our specific choice is motivated by the exceptional scattering properties of the Cr-Li system, so far overlooked, that cannot be obtained with any other atom-atom combination, and which open the way to a wealth of possibilities, going well beyond the scope of presently available systems. 
First and very importantly, the mass ratio of $^{53}$Cr and $^6$Li ($M_{Cr}/m_{Li}\sim 8.80$) is predicted to support one weakly bound (quasi bound) Cr$_2$Li trimer (Cr$_3$Li tetramer) state in the regime of strong Cr-Li repulsion \cite{Kartavtsev2007, Endo2011, Bazak2017}, which may enable the resonant tuning of genuine few-body elastic interactions, on top of the standard two-body ones. 
Second, thanks to a purely quantum interference phenomenon, three-body recombination processes to weakly bound molecular states are drastically suppressed for the Cr-Li system, when compared to any other mixture \cite{Petrov2003}, allowing for the realization of long-lived Fermi gases with strong interspecies repulsion, “immune” to the pairing instability inherent of such systems \cite{Pekker2011,Sanner2012,Amico2018}. This represents a key advantage in the experimental study of ferromagnetic phases with ultracold Fermi gases \cite{Jo2009,Sanner2012, Valtolina2017, Amico2018,Scazza2019_ii}, possibly enabling polarized domains to develop over macroscopic lengthscales, in contrast with what has been observed in homonuclear systems \cite{Amico2018,Scazza2019_ii}.
      
Besides these two main points, it is also worth noticing that fermionic $^{53}$Cr is a yet poorly explored \cite{Chicireanu2006,Naylor2015} but very interesting system on its own. Indeed, it exhibits a large magnetic dipole moment ($6\,\mu_B$, where $\mu_B$ is the Bohr’s magneton), and it is expected to feature few wide Feshbach resonances \cite{Simonipriv} which could enable precise tuning of homonuclear interactions. 
Finally, although no predictions are currently available for $^{53}$Cr-$^6$Li Feshbach resonances due to the lack of any experimental input, the rich hyperfine and Zeeman structure of these two atomic species, combined with the highly magnetic character of Cr atoms, is expected to yield rich resonance spectra, with a density about three-to-four times larger than the typical one featured by alkali mixtures \cite{Simonipriv}. In this regard, it is also important to stress that even relatively narrow Feshbach resonances could guarantee a good collisional stability for the resonantly interacting mixture \cite{Jag2016}, and they would not limit the observability of the rich few-body properties of this system, thanks to the larger Cr-Li mass ratio, compared to the K-Li one \cite{Jag2014}. 

The paper is organized as follows. In section \ref{exp} we present our new experimental setup and discuss its main features. In Section \ref{singlespecies} we report on the performance of our apparatus when operating with one single atomic species. Section \ref{doublespecies} discusses the first characterization of the cold Cr-Li mixture when the two atomic clouds are simultaneously loaded within a double-species magneto-optical trap.
Finally, in Section \ref{outlook} we summarize the outcome of our experimental studies, and discuss the prospects for the successive steps towards the ultracold regime, namely optical trapping and the implementation of evaporative and sympathetic cooling protocols.
 
\section{EXPERIMENTAL SETUP} \label{exp}
In the following, we summarize the main features of our new apparatus, whereas for a detailed description we refer the reader to Ref.\,\cite{Neri2019}. We first discuss the design strategy followed to build the vacuum and the magnetic coils setups. 
After recalling the optical transitions required for the laser cooling of the two atomic species, we then provide an overview of the specific optical setup for $^6$Li and $^{53}$Cr atoms implemented in our new machine. 

\subsection{Vacuum and magnetic coils setup}

\begin{figure*}[t]
\begin{center}
\includegraphics[width=140mm]{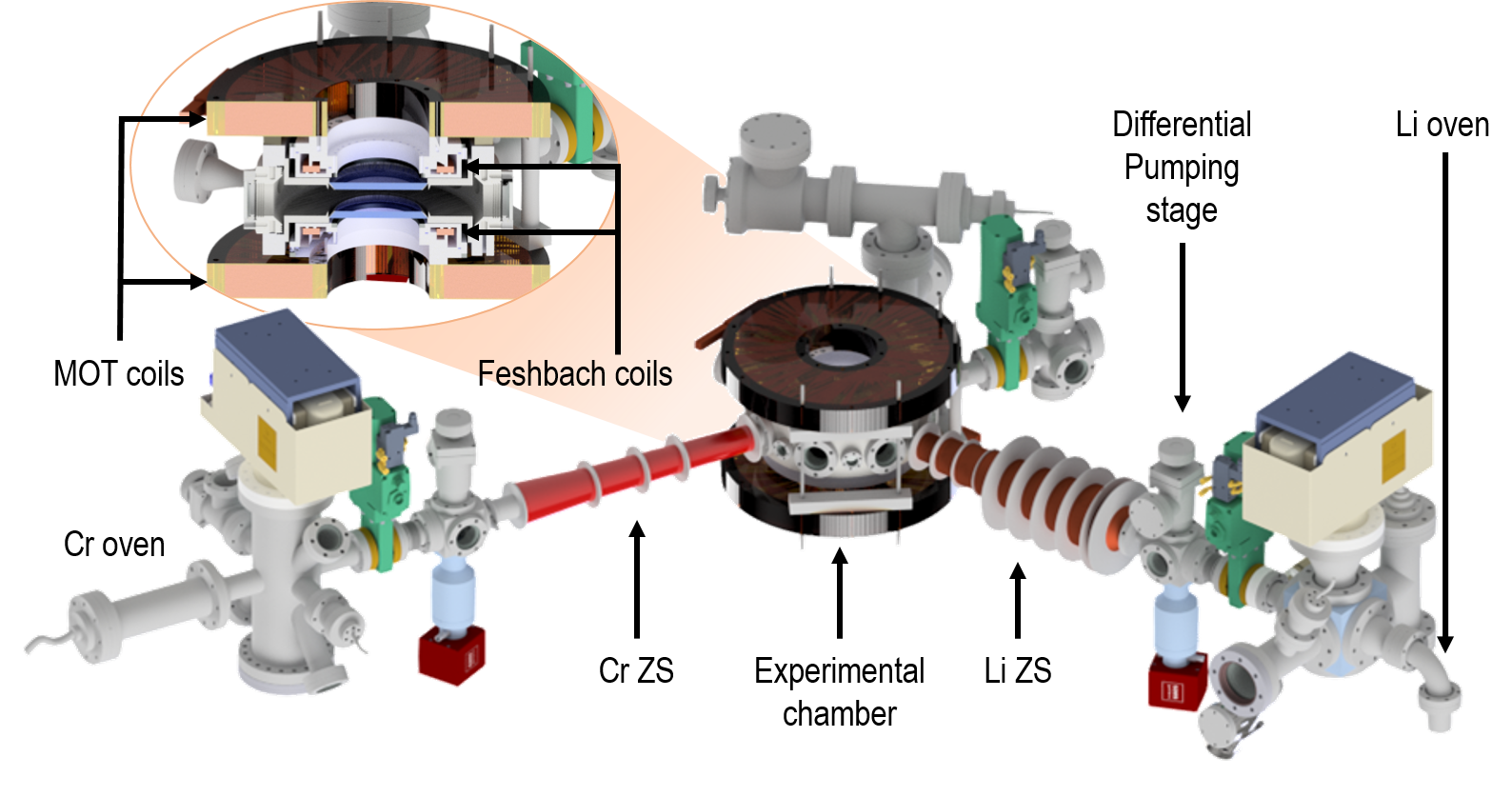}
\caption{Overview of our vacuum setup. Two independent ZS lines connect the Li and Cr effusion cells to a custom Kimball Physics spherical octagon chamber of radius $13\,$cm. In the horizontal $x-y$ plane, the cell has 8 CF40 windows, spaced out by CF16 windows. Along the vertical $z$ direction, two re-entrant CF200 windows with a wide clear aperture (radius $47.5\,$mm) ensure large optical access. Two large MOT coils, embedded within a resin structure, are fixed at the top and bottom of the science chamber, in a concentric configuration with respect to the smaller Feshbach coils, hosted within the re-entrance of the CF200 windows, see section view on the top left. The large radius of our MOT coils ensures the merging of the two ZS field profiles with the MOT quadrupole radial profile (for more details see Ref.\,\cite{Neri2019}).}
\label{Fig1}
\end{center}
\vspace*{-0pt}
\end{figure*}

Our vacuum apparatus employs two independent Zeeman slower (ZS) lines that connect the Li and Cr effusion cells to a common science chamber, see Fig.\,\ref{Fig1}. 
This choice is firstly motivated by the significantly different sublimation temperature of Li (about $410^{\circ}C$) and Cr (of  about $1500^{\circ}C$): While the former can be obtained with a custom-designed oven, for the latter we employ a commercial high-temperature effusion cell by CreaTec \cite{CrLi_sources_note}. 
Secondly, this choice is motivated by the fact that the Cr-Li chemical and collisional properties were initially unknown and, with this configuration, the two species do not affect each other up to the MOT region.

Finally, our design conveniently enables to optimize the ZS characteristics for each species independently. The obvious drawback of this geometry is that four, rather than two, viewports of the main chamber are required for the ZS vacuum lines and optical beams. In consideration of this, we employ a large radius octagon chamber, with eight perimetral $CF40$ windows spaced out by $CF16$ ones. Two large $CF200$ re-entrant windows close the chamber along the vertical direction.
Our vacuum apparatus comprises a $500\,$l/s NexTorr pump installed near the science chamber, and two differential pumping stages between the ZSs and the Cr and Li oven output, respectively. A $75\,$l/s VacIon Plus ion pump is installed at each oven. With this configuration, we ensure an ultra-high vacuum pressure below $10^{-10}\,$mbar in the MOT region.
\begin{figure*}[t!]
\begin{center}
\includegraphics[width=140mm]{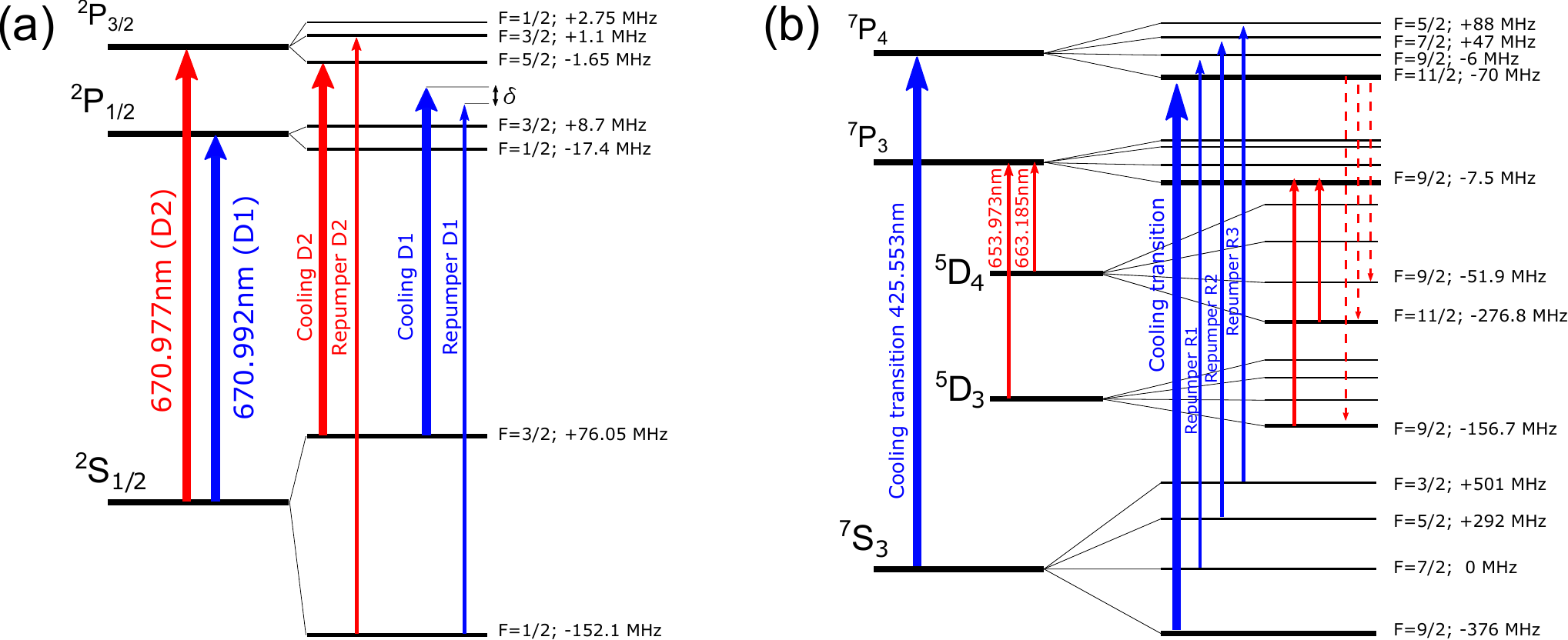}
\caption{\textbf{(a)} Sketch of the fine and hyperfine energy levels relevant for the laser cooling of $^6$Li atoms, including those exploited for the gray optical molasses based on the $D_1$ line (blue arrows). \textbf{(b)} Same as \textbf{(a)} but for $^{53}$Cr atoms. The non-zero nuclear magnetic moment ($I=3/2$) of the fermionic isotope splits each fine level into four hyperfine states. The laser cooling of $^{53}$Cr, based on the $^7S_{3} \rightarrow {^7P_{4}}$ atomic line, besides the cooling light ($F = 9/2 \rightarrow {F' = 11/2}$), also requires three blue repumpers, denoted by $R_1$, $R_2$ and $R_3$. Red dashed arrows indicate the leaks to metastable $^5D_3$ and $^5D_4$ states. Red solid arrows indicate the transitions exploited to pump $D$-state atoms back into the cooling cycle through the $^7P_{3}$ level. We determined a mass shift, on top of the hyperfine shift, of $+10.7(1.7)\,$MHz for the cooling transition of $^{53}$Cr ($F = 9/2 \rightarrow {F' = 11/2}$) referenced to the one of $^{52}$Cr ($^7S_{3}\rightarrow{^7P_{4}}$), in excellent agreement with previously reported measurements \cite{Chicireanu2006}.}
\label{Fig2}
\end{center}
\vspace*{-0pt}
\end{figure*}

The overall coil assembly is designed in such a way that the ZS axial fields merge with the MOT radial quadrupole field at the edge of the science chamber, forming two monotonically decreasing profiles. Such a non spin-flip configuration, already successfully implemented for single-species Li experiments [see e.g. Ref.\,\cite{Serwane2007}] but yet unexplored for Cr, is extremely advantageous for the fermionic isotope $^{53}$Cr, as it overcomes the difficulty connected with the existence of a “bad crossing” region along the ZS field profile near $25\,$G \cite{Chicireanu2006}: at this field, a non perfectly polarized ZS light efficiently depumps the atoms towards unaddressed hyperfine states, thus detrimentally reducing the flux of Cr atoms collected into the MOT. In our apparatus, the Zeeman slowed Cr atoms experience this field only $2\,$cm away from the quadrupole center, where they are already slow enough to be repumped and trapped by the MOT lights.
Furthermore, this design minimizes transverse-velocity induced losses at the ZS output, as it zeroes the distance that atoms, decelerated down to very low velocities, must travel before reaching the MOT region. This is a key advantage in our setup, given the large radius of $13\,$cm of the science chamber.
On the other hand, this ZS field configuration implies the use of relatively small ZS light detunings, which may sizably perturb the Cr MOT and limit its performances owing to enhanced light-assisted collisions \cite{Chicireanu2006}.
A smooth merging of the ZS and MOT field profiles at the edge of the science chamber is enabled by two large MOT coils, characterized by an inner (outer) radius of 72 mm ($160\,$mm). These are made of a square-section hollow copper wire, which allows for very efficient water cooling, and they are embedded within a custom-designed resin structure that provides a sustaining frame. 
In addition to the MOT coils, our setup is equipped with two smaller coils (inner radius of $64\,$mm; outer radius of $85\,$mm), concentric with the MOT ones, and hosted within a water-flooded toroidal PEEK case, inserted within the re-entrance of the $CF200$ viewports, see section view on the top left in Fig.\,\ref{Fig1}. 
This second pair of coils is primarily designed to create homogeneous Feshbach fields up to $1000\,$G with less than $200\,$A current. Moreover, when switched to anti-Helmoltz configuration, the Feshbach coils can provide a quadrupole field with fast switch-off timescales ($<1\,$ms), thanks to their small inductance.
Three additional pairs of coils placed along three orthogonal directions enable to create fields of a few Gauss, to finely adjust the quadrupole position and to compensate for spurious offset fields.

\begin{figure}[t!]
\begin{center}
\includegraphics[width=85mm]{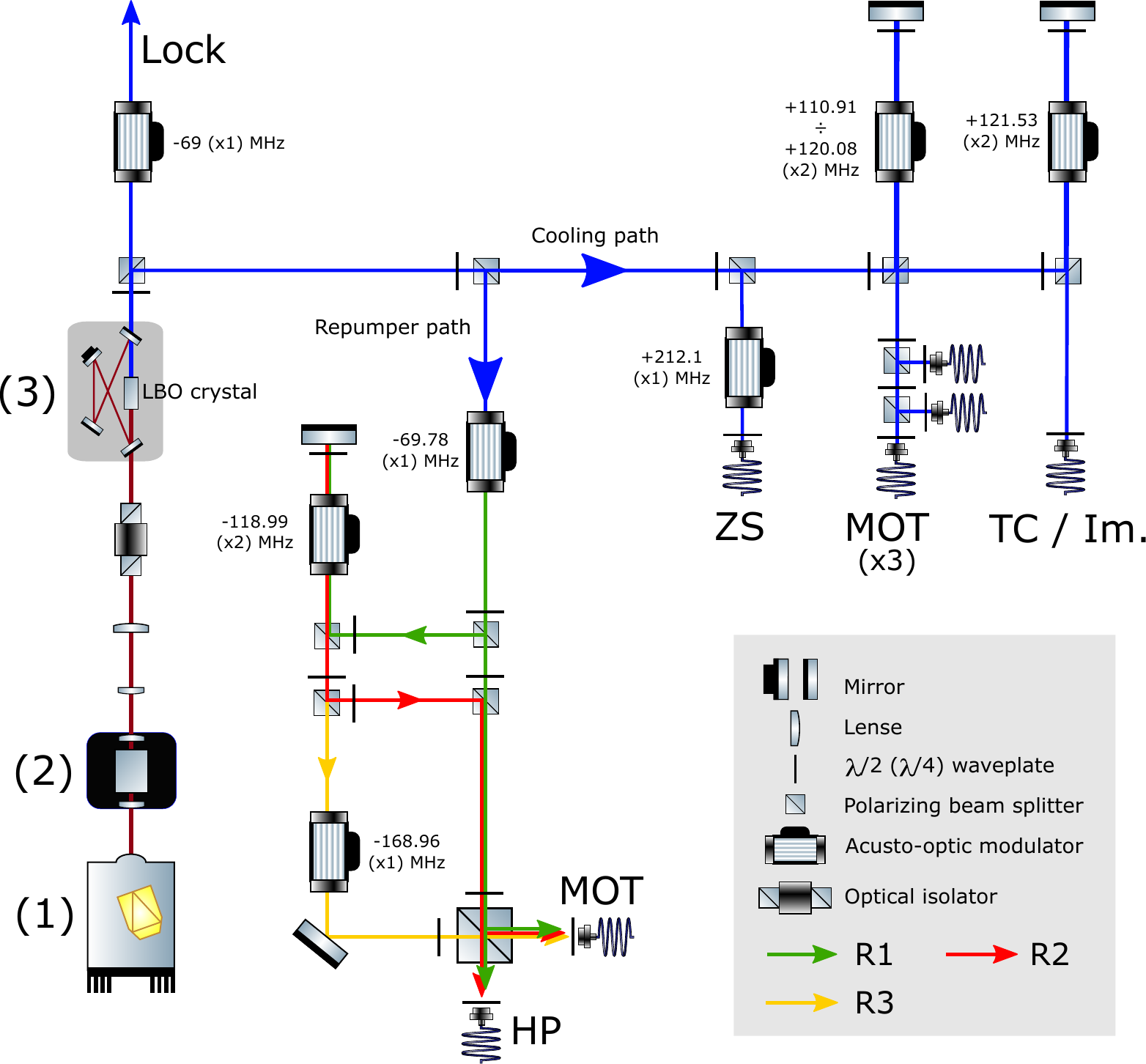}
\caption{Simplified view of the optical setup employed to obtain the blue lights required for the laser cooling of $^{53}$Cr. The $425.5\,$nm light is produced from a $851\,$nm diode laser (1) through a frequency-doubling bow-tie cavity (3). A tapered amplifier chip hosted on a home-made mount (2) amplifies the infrared light, while a system of spherical and cylindrical lenses collimates the beam, see Ref.\,\cite{Neri2019} for details. A pickup beam at the cavity output is sent to the locking setup. An AOM along its path sets the light to be blue-detuned from the $^{52}$Cr $^7S_{3} \rightarrow {^7P_{4}}$ transition by $69\,$MHz. The remaining light is split into the cooling and repumper beam paths. Along the first one, one single-pass and two double-pass AOMs produce the ZS, MOT and TC beams, respectively. From the TC beam, the imaging beam is also derived. Along the repumper path, a series of three AOMs produces the $R_1$, $R_2$ and $R_3$ lights. From them, we produce the repumper beam for the MOT, containing all three lights, and the HP beam, containing only $R_1$ and $R_2$.}
\label{FigCROptSetup}
\end{center}
\end{figure}
\subsection{Lithium Laser Setup}

The atomic transitions employed for the laser cooling of Li atoms are schematically shown in Fig.\,\ref{Fig2}(a). The Li MOT is based on the $D_2$ ($^2S_{1/2}\rightarrow{^2P_{3/2}}$) atomic line at $671\,$nm, featuring a natural linewidth $\Gamma_{Li}/2\pi=5.87\,$MHz, and a saturation intensity $I_{S,Li}=2.54\,$mW/cm$^2$. The $D_2$ cooling and repumper lights, detuned by $228\,$MHz  from each other, address the $F=3/2\rightarrow {F' =5/2}$ and the $F =1/2 \rightarrow {F' = 3/2}$ transition, respectively. Two additional laser lights addressing the hyperfine transitions associated with the $D_1$ ($^2S_{1/2}\rightarrow{^2P_{1/2}}$) atomic line, red-detuned by about $10\,$GHz from the $D_2$ line, are employed for a sub-Doppler cooling stage based on gray optical molasses \cite{Burchianti2014}.

Our optical setup is essentially analogous to the one previously developed in the Lithium lab at LENS \cite{ValtolinaPhD}. The $D_2$ and the $D_1$ lights are provided by two separate master laser sources, and independently controlled by two acousto-optic modulators (AOMs) which act as fast switches. The two AOM output beams are immediately recombined on a common optical path. They are split into cooling and repumper lights, whose frequency and power are independently adjusted by means of two AOMs, and injected into two commercial tapered amplifiers, each one delivering about $300\,$mW. The cooling and repumper lights are then recombined on a $50:50$ non-polarizing beamsplitter, and sent to the MOT and ZS/imaging beam paths. Three fibers bring the MOT light to the science chamber: Three mutually orthogonal retroreflected beams with a $1/e^2$ radius of about $5.0\,$mm and total peak intensity of about $29\,I_{S,Li}$ realize the MOT configuration.
The ZS/imaging path includes an additional single-pass AOM, and its frequency is toggled to inject the light into either the ZS or the imaging fiber.
At the entrance of the vacuum chamber, the ZS beam, slightly focused, features a $1/e^2$ radius of about $2.2\,$mm, with total peak intensity of about $200\,I_{S,Li}$. 

\subsection{Chromium Laser Setup}
The optical setup required for the laser cooling of fermionic $^{53}$Cr atoms is rather complex, owing to the presence of metastable $D$-states into which atoms can leak from the cooling cycle, and to a rich hyperfine structure [see Fig.\,\ref{Fig2}(b)]. Moreover, a transverse cooling (TC) stage at the oven chamber is known to be crucial \cite{Chicireanu2006}  to increase the atomic beam collimation, and thus the flux of Zeeman slowed atoms at the MOT region.
The laser cooling scheme is based on the $^7S_{3} \rightarrow{^7P_{4}}$ atomic line at $425.5\,$nm, with natural linewidth $\Gamma_{Cr}/2\pi=5.06\,$MHz and saturation intensity $I_{S,Cr} =8.52\,$mW/cm$^2$. While for bosonic Cr isotopes one single frequency is needed to operate the MOT, the four hyperfine sub-levels of the $^{53}$Cr fine states require additional repumper beams, besides the $F = 9/2 \rightarrow {F' = 11/2}$ cooling light. Up to three blue repumpers, denoted by $R1$, $R2$ and $R3$ in Fig.\,\ref{Fig2}(b), are in principle needed to bring excited $P$-state atoms back into the cooling cycle. Moreover, even with all blue repumpers on, the MOT transition remains slightly leaky, since optically excited atoms can decay from the $^7P_4$ state onto the underlying $^5D_{3,4}$ metastable states. Therefore, additional “red” repumpers are needed to fully close the cooling cycle \cite{Chicireanu2006}. These  lights, at a wavelength of $654\,$nm and $663\,$nm, pump atoms from the metastable $^5D_3$ and $^5D_4$ states, respectively, back to the ground state via the $^7P_3$ level, see red solid arrows in Fig.\,\ref{Fig2}(b). 

In the following, we provide a detailed description of our Cr optical setup and its distinctive features, as compared to the one employed by the only other group that has successfully laser cooled $^{53}$Cr thus far \cite{Chicireanu2006}.
The setup is schematically shown in Fig.\,\ref{FigCROptSetup}. The blue light is generated by frequency-doubling a high-power laser source at $851\,$nm using a non-linear lithium triborate (LBO) crystal within a custom-designed bow-tie optical cavity \cite{Neri2019}. Up to $3\,$W of $851\,$nm light are obtained by seeding a tapered amplifier with a commercial diode laser. With an overall conversion efficiency of $30$\%, the doubling cavity, injected with $2\,$W of infrared light, delivers about $600\,$mW at $425.5\,$nm.
$10\,$mW of this light are used to lock the master laser, via a standard saturation spectroscopy scheme on a hollow cathode lamp, such that the blue light is resonant with the $^{52}$Cr $^7S_{3} \rightarrow {^7P_{4}}$ transition. The remaining $425.5\,$nm light is split into two main optical paths for the cooling and the repumpers, respectively.

The cooling beam is further split into ZS, MOT and TC paths, see Fig.\,\ref{FigCROptSetup}. Similarly to the case of Li, our MOT light configuration consists of three mutually orthogonal retroreflected beams. These are characterized by a $1/e^2$ radius of $3.0\,$mm and an average peak intensity of $2.3\,I_{S,Cr}$.
The TC light is sent near the output of the Cr oven, where it is split into two retroreflected orthogonal beams, perpendicular to the ZS direction. In order to maximize the interaction time of the atoms with the cooling light, the TC beams have an elliptical shape of $1 \div 3$ aspect ratio, with the large waist of $4.1\,$mm oriented along the ZS direction. A total peak intensity of $20\,I_{S,Cr}$ is employed for the two TC beams, and their frequency is set close to resonance with the $F=9/2 \rightarrow {F'=11/2}$ transition. A pickup of the TC light is coupled into a fiber, delivering $200\,\mu$W, employed for absorption imaging.
The ZS beam, characterized by a waist of $3.1\,$mm at the MOT region, is slightly focusing and it is directed towards the oven output by means of an in-vacuum mirror. The light is circularly polarized as to address mostly $\sigma_+$ transitions, with an intensity of $32\,I_{S,Cr}$ at the MOT region.

From the repumper path we obtain the $R_1$, $R_2$ and $R_3$ lights, indicated in Fig.\,\ref{FigCROptSetup} by green, red and yellow arrows, respectively. Their frequencies are adjusted by means of the three AOMs, and they are kept constant throughout the experimental cycle. 
By combining all repumper lights at a polarizing beam splitter we obtain two beams. The first one, containing all three lights, is sent to the science chamber to create a single retroreflected beam, crossing the MOT region with an angle of $22.5$ degrees with respect both to one of the horizontal MOT directions and to the ZS axis. 
It is characterized by a $1/e^2$ radius of $3.4\,$mm and total peak intensities of $12$ ($R_1$), $6$ ($R_2$) and $6\,I_{S,Cr}$ ($R_3$), respectively.
The second beam, containing only $R_1$ and $R_2$, is sent to the Cr oven output. It is exploited to increase the atomic population of the lowest hyperfine $F=9/2$ manifold, and thus the flux of Zeeman slowed atoms. This “hyperfine pumping” (HP) beam has a $1/e^2$ radius of about $2.2\,$mm ($0.7\,$mm) along the longitudinal (transverse) direction of the atomic beam, and a total peak intensity of $90\,I_{S,Cr}$, equally distributed between $R_1$ and $R_2$. 

Finally, besides the blue lights, our setup also comprises two “red” repumpers, realized by two commercial diode lasers at $654\,$nm and $663\,$nm. Their proximity to the Li atomic reference at $671\,$nm enables a simple and cost-effective locking scheme of their frequency, based on a commercial Fabry-Pérot resonator as transfer cavity [see Ref.\,\cite{Neri2019} for details].
 
\section{EXPERIMENTAL RESULTS: SINGLE SPECIES} \label{singlespecies}
In the following, we discuss the experimental procedures exploited to produce individual clouds of cold Li and Cr atoms, and summarize the performance of our machine under optimized conditions for single species operation.

\subsection{Lithium}
Our strategy to laser cool and manipulate $^6$Li atoms is essentially analogous to the one already developed in the Lithium lab at LENS \cite{Burchianti2014}, the only difference being the non spin-flip configuration of the ZS employed in our setup. 
Optimization of the experimental cycle is obtained by monitoring the MOT fluorescence signal on a calibrated photo-detector, and by acquiring time-of-flight (TOF) absorption images of the Li clouds on a CCD camera, from which we deduce the gas temperature.
\begin{figure}[t!]
\begin{center}
\includegraphics[width= 8.0cm]{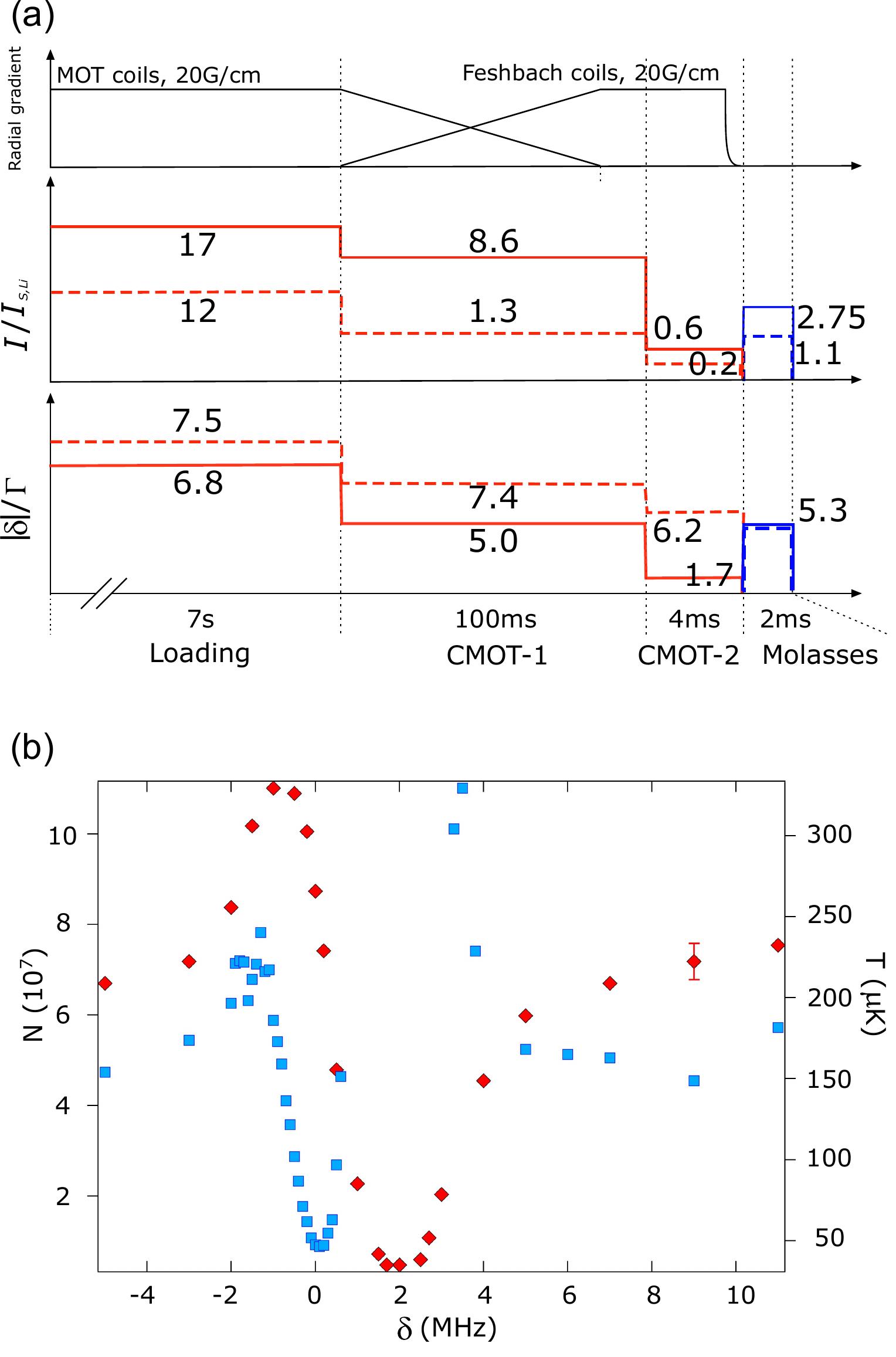}
\caption{\textbf{(a)} Sketch of the experimental routine employed to produce cold clouds of $^6$Li atoms in our experiment. For each stage, red (blue) solid and dashed lines indicate the parameters exploited for $D_2$ ($D_1$) cooling and repumper lights, respectively. The transfer from the MOT coils to the Feshbach coils quadrupole is also shown by the black solid lines. \textbf{(b)} Temperature (right axis, blue squares) and atom number (left axis, red diamonds) as a function of the relative detuning $\delta$ of the $D_1$ repumper and cooling lights, recorded after $2\,$ms of gray molasses. At the Raman condition, $\delta=0$, a minimum temperature of $45(5)\,\mu$K is obtained, with a cooled fraction $N/N_0 \sim 75$\%. For this measurement, the total cooling and repumper peak intensities were fixed to the optimum values $I_{cool}=2.75\,I_{S,Li}$ and $I_{rep}\sim 0.4\,I_{cool}$, respectively.}
\label{FigLi}
\end{center}
\end{figure}
Fig.\,\ref{FigLi}(a) schematically illustrates the sequence employed in the experiment, which consists of the initial MOT loading, followed by two compressed-MOT (CMOT) phases, all operating on the $D_2$ transition. These are followed by a gray optical molasses stage based on the $D_1$ line. 
Table \ref{TabLi} summarizes the corresponding atom number, peak density and temperature of the cloud, measured at each step of the routine after $7\,$s loading and with the following experimental parameters: oven temperature set to $410^{\circ}\,$C, ZS total power (detuning) of $40\,$mW ($-21.3\,\Gamma$), total MOT peak intensity of $29\,I_{S,Li}$. 
During the loading stage, similarly to Ref.\,\cite{Burchianti2014}, we keep the detuning of the MOT cooling (repumper) light at about $-6.8\,\Gamma$ ($-7.5\,\Gamma$), to maximize the number of trapped atoms while limiting the gas temperature to about $2.5\,$ mK.
\begin{table}[]
\begin{tabular}{|l|l|l|l|l|}
\hline
\multicolumn{1}{|c|}{\textbf{Stage}} & \multicolumn{1}{c|}{\textbf{MOT}} & \multicolumn{1}{c|}{\textbf{CMOT$_1$}} & \multicolumn{1}{c|}{\textbf{CMOT$_2$}} & \multicolumn{1}{c|}{\textbf{$D_1$}} \\ \hline \hline
Atom number \small{($10^8$)}& 2.0(2)                                  & 2.0(2)                                    & 1.7(2)                                  & 1.27(12)                             \\ \hline
Peak density \small{($10^{9} cm^{-3}$)}& 0.64(12)                               &4(1)        &21(4)        &24(6)                 \\ \hline
Temperature  \small{($\mu$K)}&{2500}         &{900(100)}                  &500(50)                  &{45(5)}                 \\ \hline
\end{tabular}
\caption{Typical atom number, peak density and temperature of our $^6$Li cloud recorded after each stage of the experimental cycle, for a $7\,$s loading time and the experimental settings detailed in the text.}
\label{TabLi}
\end{table}
We then turn off both the ZS field and light, and we reduce the temperature of the Li cloud by performing two successive stages of CMOT, during which we decrease both the detuning and intensity of the MOT beams, see Fig.\,\ref{FigLi}(a). 
A first CMOT stage, lasting for $100\,$ms, enables to lower the temperature to about $0.9(1)\,$mK. Simultaneously, we adiabatically transfer the cloud from the MOT to the Feshbach coils quadrupole. Owing to their much smaller inductance, indeed, the Feshbach coils allow for a fast switch-off time of about $500\,\mu$s at typical operating currents. 
A second CMOT stage, lasting about $4\,$ms, further cools the Li MOT below $500\,\mu$K, with atom losses below $20$\%. 
The average $1/e$ MOT radius drops from the initial $3.8(3)\,$mm value down to $1.20(14)\,$mm and, correspondingly, the peak density increases from $6.4(1.2)\,10^{8}\,$atoms/cm$^3$ to $2.1(4)\,10^{10}\,$ atoms/cm$^3$.

We then turn off the $D_2$ MOT light and quadrupole field while turning on the $D_1$ cooling and repumper lights, both blue-detuned with respect to their corresponding resonances.
We have tested the performance of $D_1$ molasses upon varying the light parameters, by monitoring the temperature and the atom number through TOF absorption imaging, resonant with the $D_2$ $F=1/2 \rightarrow {F’=3/2}$ transition. 
Fig.\,\ref{FigLi}(b) shows the evolution of temperature and atom number, after 2 ms of gray molasses, as a function of the relative detuning, $\delta = \delta_1 - \delta_2$. 
Here, $\delta_1$ and $\delta_2$ are the detuning of the $D_1$ repumper and cooling lights from the $F=1/2 \rightarrow {F'=3/2}$ and $F=3/2 \rightarrow {F'=3/2}$ transitions, respectively [see Fig.\,\ref{Fig2}(a)]. 
The cloud temperature as a function of $\delta$ exhibits an asymmetric Fano profile with sub-natural width centered around the Raman resonance ($\delta=0$). This trend represents the distinctive feature of such a sub-Doppler cooling technique, that arises from the Sisyphus effect on the blue $F \rightarrow {F' = F}$ transition combined with the emergence of a coherent dark state, see e.g. Ref.\,\cite{Grier2013} for details. 
In particular, at the Raman condition, the cloud temperature is found to reach a minimum value of $T=45(5)\, \mu$K, with a cooled fraction $N/N_0$ of about $75$\%, in excellent agreement with what was previously observed in the Lithium lab at LENS \cite{Burchianti2014}. 

\subsection{Chromium}

In contrast with the well-established Li system, fermionic Cr is still a relatively poorly explored species in the cold regime. Thus far, indeed, only one group worldwide successfully laser-cooled $^{53}$Cr atoms \cite{Chicireanu2006} and, more recently, brought it to quantum degeneracy by means of sympathetic cooling with the most abundant bosonic isotope $^{52}$Cr \cite{Naylor2015}.
For this reason, in the following we provide a detailed description of our experimental protocol, which enabled to realize samples of almost $10^7$ $^{53}$Cr atoms and temperatures below 150 $\mu$K, discussing the similarities and differences between our working conditions and the Paris ones \cite{Chicireanu2006}. 
A few peculiar features make laser cooling of Cr isotopes rather challenging. First, the anomalously large light-assisted collision rates of this species limit the $^{52}$Cr ($^{53}$Cr) steady-state MOT population to a few $10^7$ ($10^5$) [see \cite{Chicireanu2006} and references therein]. Secondly, as discussed in  Section \ref{exp}, the radiative decay of $P$-state atoms towards metastable $D$-states requires the implementation of a laser setup significantly more complex than for alkali or lanthanide atoms. Thirdly, when the fermionic $^{53}$Cr isotope is considered, its rich hyperfine structure and the relatively small natural abundance strongly decreases the flux of Zeeman-slowed atoms that can be experimentally  achieved. In fact, while relatively high loading rates of $\Gamma_L \sim 3\, 10^8$ atoms/s can be obtained for $^{52}$Cr, $^{53}$Cr MOTs usually exhibit $\Gamma_L$ of the order of $10^6$ atoms/s \cite{Chicireanu2006}. Such a reduction of more than two orders of magnitude arises from the combination of a $9$-fold lower natural abundance, together with the fact that only one among the 28 Zeeman sublevels of the electronic ground state, specifically the $|F,m_F\rangle {=} |9/2,+9/2\rangle$ state, can be decelerated within the ZS.

\begin{figure}[t!]
\begin{center}
\includegraphics[width= 8cm]{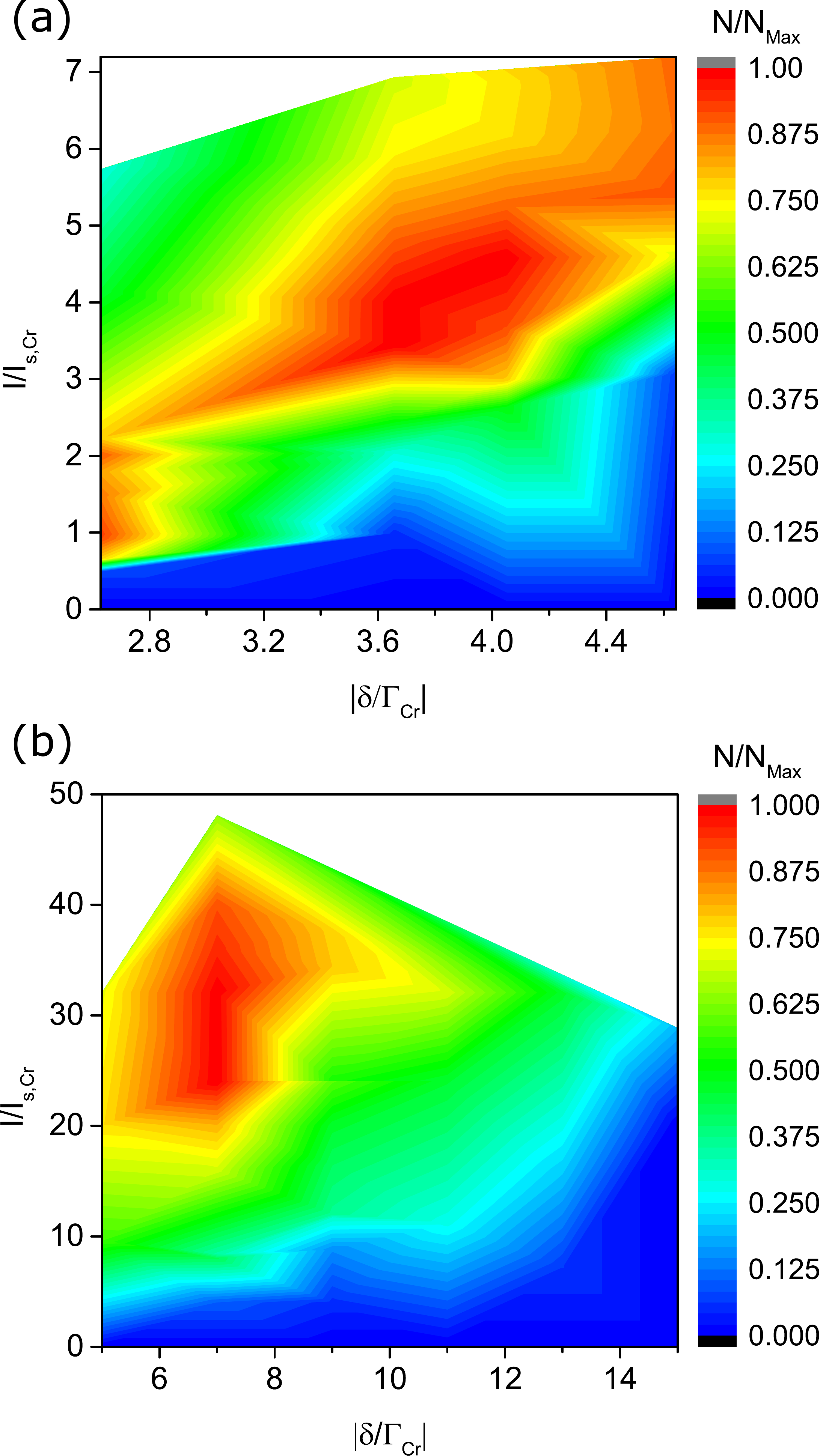}
\caption{\textbf{(a)} Steady-state MOT population, rescaled to its optimum value, as a function of both the normalized single-beam peak intensity $I/I_{S,Cr}$ and the detuning $\delta/\Gamma_{Cr}$ of the cooling light, see color code on the right. This measurement employs a radial MOT gradient of $12.6\,$G/cm, and a ZS beam intensity (detuning) of $32\,I_{S,Cr}$ ($-7\,\Gamma_{Cr}$). \textbf{(b)} Same as panel \textbf{(a)} but for the ZS cooling beam parameters. The given ZS beam intensity corresponds to the one estimated at the MOT region. For this measurement, the single MOT beam peak intensity (detuning) is set to $3.5\,I/I_{S,Cr}$ ($-3.5\,\Gamma_{Cr}$).}
\label{FigCr}
\end{center}
\end{figure}

\begin{figure}[t!]
\begin{center}
\includegraphics[width= 6.5cm]{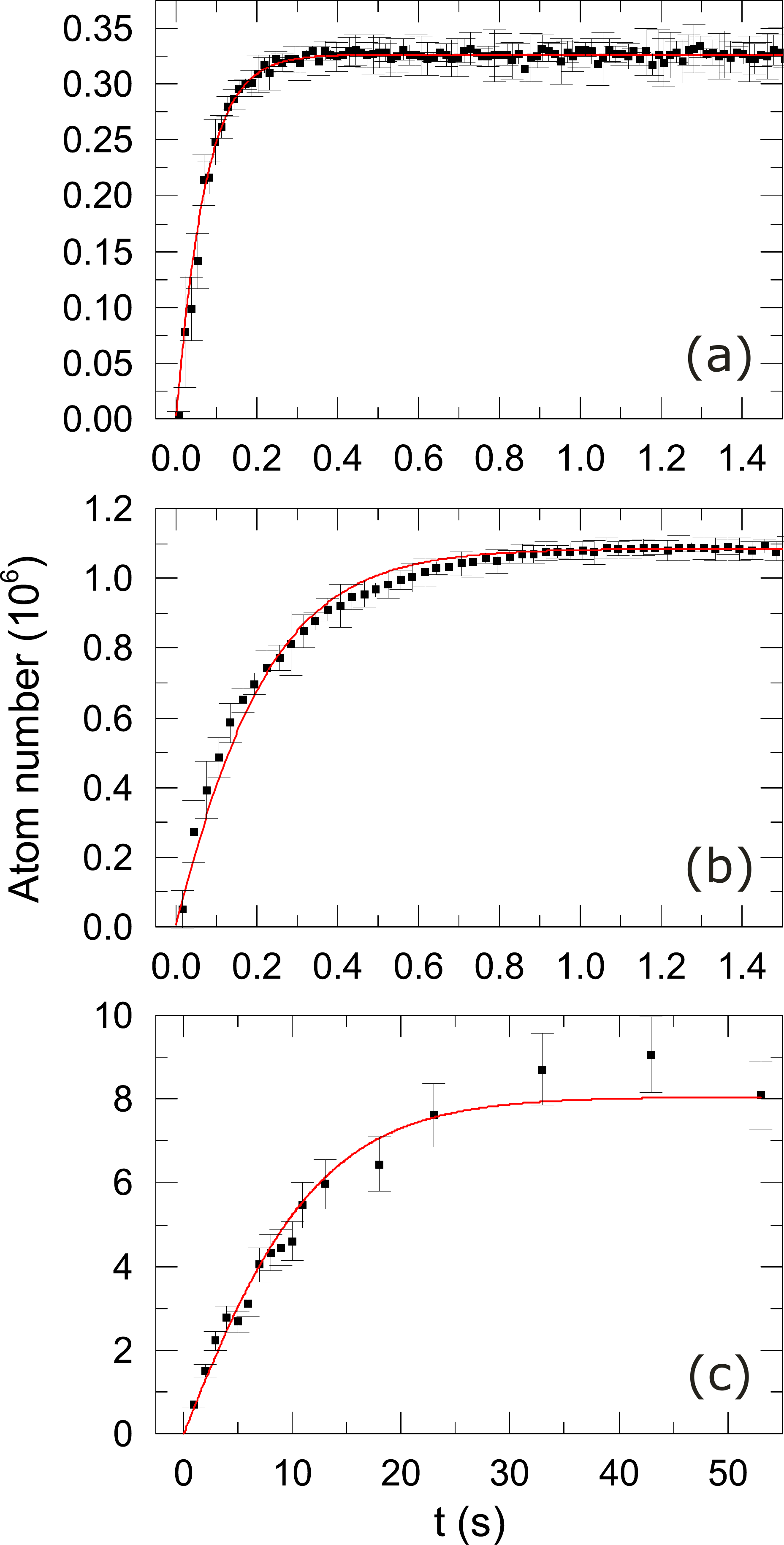}
\caption{\textbf{(a)} Loading curve of a $^{53}$Cr MOT without red repumpers. Data are obtained by recording the MOT fluorescence as a function of the loading time on a CMOS camera with a $60(10)\,$Hz frame rate. The y-axis is calibrated by independently measuring the steady-state atom number via absorption imaging. \textbf{(b)} Same as panel \textbf{(a)}, but with the red repumpers permanently on. \textbf{(c)} Loading curve of $D$-state atoms accumulated in the magnetic quadrupole and recaptured within the MOT after application of a 30-ms-long pulse of the red repumpers. The atom number is in this case monitored by means of absorption imaging. Red solid curves are the best fit to the models described in the main text.}
\label{CrLoading}
\end{center}
\end{figure}

In our experiment, we have first maximized the number of collected $^{53}$Cr atoms by thoroughly scanning all MOT, TC, HP and ZS parameters. 
Examples of such a characterization are presented in the contour plots of Fig.\,\ref{FigCr}. There we show the dependence of the steady-state MOT atom number on both the peak intensity and detuning of the cooling light MOT [Fig.\,\ref{FigCr}(a)] and ZS [Fig.\,\ref{FigCr}(b)] beams. 
The observed trend nicely matches the results of previous studies \cite{Chicireanu2006}. In particular, an almost constant maximum value of the MOT population is obtained for detunings $3.4\lesssim|\delta|/\Gamma_{Cr}\lesssim 4.2$ at single-beam intensities $3 \lesssim I/I_{S,Cr} \lesssim 5$. The decrease of the MOT performances for $|\delta|< 3.4\,\Gamma_{Cr}$ can be  understood by considering that smaller detunings cause an increase of both the atomic density and the light-assisted collision rate \cite{ChicireanuPhD}. 
Moving to the opposite limit, we observe a decrease in the atom number for $|\delta| >5 \, \Gamma_{Cr}$ which cannot be compensated any more upon increasing the beams intensity. From this we deduce that the MOT capture volume is limited in this regime by our MOT beams size. Under the working conditions yielding the largest atom number, we obtain samples of up to $3.5(2)\,10^5$ atoms. 
By fitting the initial slope of the measured loading curves [see example in Fig.\,\ref{CrLoading}(a)], characterized by a typical $1/e$ time of about 65(5) ms, we infer $^{53}$Cr loading rates of about $\Gamma_L \sim 4-6 \, 10^6\,$atoms/s for an oven temperature of $1500^{\circ}$C. These values are comparable with, and even slightly larger than, the ones previously reported by the Paris group \cite{Chicireanu2006,ChicireanuPhD} under similar MOT gradient and cooling light  parameters. 

\begin{table}[t!]
\begin{tabular}{|p{1cm}|p{1cm}|p{1cm}|p{1cm}|p{1cm}|p{1cm}|}
\hline
\multicolumn{1}{|c|}{\textbf{\small{No R$_1$}}} & \multicolumn{1}{c|}{\textbf{\small{No R$_{2,3}$}}} & \multicolumn{1}{c|}{\textbf{\small{No R$_2$}}} & \multicolumn{1}{c|}{\textbf{\small{No R$_3$}}} & \multicolumn{1}{c|}{\textbf{\small{No TC}}} & \multicolumn{1}{c|}{\textbf{\small{No  HP}}}\\ \hline \hline
0&0&0.27(3)&0.54(4)&0.21(6)&0.42(5)\\ \hline 
\end{tabular}

\begin{tabular}{|p{2.35cm}|p{2.35cm}|p{2.35cm}|}
\hline
\multicolumn{1}{|c|}{\textbf{Light}} & \multicolumn{1}{c|}{\textbf{$I/I_{S,Cr}$}} & \multicolumn{1}{c|}{\textbf{$|\delta|/\Gamma_{Cr}$}}\\ \hline \hline
ZS cooling & 32     & 6.94  \\ \hline
TC cooling & 20    & 0.86 \\ \hline
MOT cooling (per beam) &2.3 & 4.02\\ \hline
HP R$_1$  & 45    & 0.94\\ \hline
HP R$_2$  & 45     & 0.43 \\ \hline
MOT R$_1$  &12.4 & 0.94 \\ \hline
MOT R$_2$  &5.9 & 0.43\\ \hline
MOT R$_3$  &6.7 & 0.63 \\ \hline
\end{tabular}
\caption{\textbf{Top panel.} Effect of the different beams on the relative  $^{53}$Cr atom number in the steady-state MOT without red repumpers. The same factors apply to the population of magnetically-trapped D-state atoms.
\textbf{Bottom panel.} Light parameters for an optimum MOT loading at a ZS current of $14.5\,$A and a MOT current of $25\,$A, corresponding to a radial gradient of $12.6\,$G/cm. The detuning of each light is measured from the corresponding relevant transition.}
\label{Table2}
\end{table}

Despite these similarities with the Paris experiment, we emphasize that in our case this MOT performance is obtained under very different conditions of ZS and repumping light parameters.
First of all, in our setup no repumping lights are required in the ZS beam to reach the aforementioned loading rate. This is due to our non spin-flip configuration, where the “bad crossing” near $25$ G along the ZS field profile is produced very close to the MOT region, as detailed in Section \ref{exp}. Secondly, our ZS optimally works at a much smaller light detuning, of about $-7\,\Gamma_{Cr}$ [see Fig.\,\ref{FigCr}(b)], corresponding to a ZS exit velocity at zero magnetic field of $15\,$m/s. Such a low value is not constrained by the MOT parameters of our setup, and it appears to be inherently related to the hyperfine structure of $^{53}$Cr. In fact, although not discussed in the present paper, while investigating the bosonic $^{52}$Cr isotope \cite{Neri2019}, our MOT could capture atoms traveling at a threefold higher velocity. In the case of fermionic $^{53}$Cr, atoms exit our ZS in the $|9/2,+9/2\rangle$ Zeeman level of the lowest hyperfine manifold and, after crossing the quadrupole center, they must scatter many $\sigma_-$ photons to be captured in the MOT. 
During this final deceleration stage, transitions to hyperfine levels other than $F'=11/2$ are not forbidden, and they are even enhanced by the residual Doppler shift, making the Zeeman slowed atoms likely to fall into $F \neq 9/2$ states, out of the cooling cycle. 
This presumably does not happen with the spin-flip ZS of the Paris setup, which mostly delivers atoms in the $|9/2, -9/2\rangle$ state, for which undesired transitions are suppressed by the MOT light polarization.
As a result, all three repumpers are essential to optimally capture the $^{53}$Cr Zeeman slowed atoms within our MOT, as summarized in the top row of Table \ref{Table2}. 

Besides $R_1$, we need at least a second repumper to form the $^{53}$Cr MOT, and the removal of the $R_2$ ($R_3$) light causes a three- (two-) fold reduction of the collected atoms. This starkly contrasts with what observed in the Paris experiment \cite{Chicireanu2006}, where removing $R_2$ caused only a $30$\% reduction of the MOT atom number, and the effect of $R_3$ was inferred to be negligible. 
We ascribe this discrepancy to the opposite polarization of the atoms exiting the ZS in the two setups. In particular, even a relatively small residual Doppler shift makes the $R_1$ repumper light resonant with the undesired $F=7/2\rightarrow {F'=7/2}$ transition, further increasing the leak towards higher-lying hyperfine manifolds, and thus making the additional repumpers essential for operating our MOT. 
This interpretation is supported by noticing that $R_2$ and $R_3$ are irrelevant for the blue MOT, when this is seeded by cold $D$-state atoms trapped in the magnetic quadrupole [see discussion below], and characterized by thermal velocities more than $50$ times lower than our ZS exit velocity.
Differently from the Paris experiment, we shine the repumper lights onto the MOT via a single retroreflected beam. We have verified that this setup yields a repumping efficiency  equivalent to that of a three-beams configuration. Furthermore, with such a scheme one has the advantage that $R_1$, essentially coincident with the $^{52}$Cr cooling light, does not create an undesired bosonic MOT on top of the fermionic one. 

Another relevant difference in our experimental strategy with respect to the one exploited by the Paris group is the implementation of the HP beam, combined with the TC stage, at the Cr oven output and in absence of external magnetic fields. 
The increment in the MOT atom number produced by the HP and TC stages is relevant, yielding about a two-fold and five-fold increase of the atomic flux, respectively [see top row of Table \ref{Table2}]. 
We have found that both the HP efficiency and the MOT performances are essentially constant within the range of detunings $-1.5\lesssim \delta/\Gamma_{Cr}\lesssim0$ for the three blue repumpers. This allows us to derive the repumper lights for the HP and the MOT beams from the same AOM setup [see Fig.\,\ref{FigCROptSetup}].
By contrast, the final temperature of the $^{53}$Cr cloud can be strongly reduced by employing slightly red-detuned repumpers during a final CMOT stage [see discussion at the end of this section]. As such, the detunings of $R_1$, $R_2$ and $R_3$ are fixed, throughout the experimental cycle, to the small values listed in the bottom part of Table \ref{Table2}, which summarizes our optimum light parameters. 

We now move to discuss how the collected atom number can be substantially increased by employing two additional red repumper lights. Indeed, as already discussed in the previous sections, the MOT transition is not perfectly closed even when all blue repumpers are exploited, due to the leak from the excited $^7P_4$ state towards metastable $^5D$ states, with an overall rate of a few hundred Hz \cite{Chicireanu2006,ChicireanuPhD}. 
The red repumping lights at $663\,$nm and $654\,$nm, addressing the $^5D_{4}\rightarrow {^7P_3}$ and the $^5D_{3}\rightarrow {^7P_3}$ transitions, respectively, bring $D$-state atoms back into the ground state via the $^7P_3$ state, thus closing the cooling cycle. 
In principle, fermionic Cr, contrarily to the bosonic isotopes lacking hyperfine structure, would require three, rather than two, distinct red repumpers, as atoms excited to the $F=11/2$ hyperfine manifold of the $^7P_4$ level, can decay onto the $F=9/2$ manifold of the $^5D_{3}$ state, and onto the $F=11/2$ and $F=9/2$ manifolds of the $^5D_{4}$ one [see Fig.\,\ref{Fig2}(b)].

We have experimentally located the three red repumper resonances by scanning the frequency of each of the two lasers, continuously shined onto the atomic cloud. Hitting one resonance is signaled by a sudden increase of the MOT fluorescence, as the red repumper light prevents the accumulation of atoms into the dark $D$-states.
The results of this spectroscopic study are in excellent agreement with the findings of the Paris group \cite{ChicireanuPhD}.
Although the shift of about $250\,$MHz between the two $^5D_{4}\rightarrow {^7P_3}$ transitions could be easily bridged by one additional AOM, for simplicity in this work we have not employed the $663\,$nm repumper addressing the $F=9/2$ manifold, as this gives the smallest atom number increase. 

The impact of the red repumpers on the MOT properties can be noticed by comparing the loading curves recorded with only the blue light on, see Fig.\,\ref{CrLoading}(a), and with the addition of the red lights, constantly illuminating the cold sample, see Fig.\,\ref{CrLoading}(b). 
One can see how in the latter case the steady-state MOT population features a three-fold increase, relative to the former one. Parallel to this, while the loading rate $\Gamma_L$ remains unchanged within our accuracy, the $1/e$ characteristic time of the MOT loading curve increases from about $70\,$ms to $200\,$ms.
These results, fully consistent with previous observations \cite{Chicireanu2006}, can be  understood by considering that the time evolution of the MOT population, under fixed loading conditions, is  determined by the concurrent action of radiative decay and light-assisted collisions:
\begin{eqnarray}
N_{M}^{'}(t) =\Gamma_{L}-\gamma_D N_{M}(t)-\beta^{'}_{MM}N_{M}^2(t).
\label{MOTeq}
\end{eqnarray}
Here $\gamma_D=\Pi_P\gamma_{P\rightarrow {D}}$ is the decay rate towards the $D$-states, weighted by the relative population of $P$-state atoms $\Pi_P$, which in turn depends on the photon scattering rate within the MOT. 
The two-body loss term is $\beta^{'}_{MM}=\frac{\beta_{MM}}{V_{MM}}$, where $\beta_{MM}$ is the rate coefficient per unit volume for light-assisted collisions, and $V_{MM}= N_M^2/(\int d\textbf{r} \,n_M^2(\textbf{r}))$ is the MOT collisional volume. This is assumed to be constant in time, an approximation fully validated by early experimental studies \cite{Bradley2000}. We remark that Eq.\,\eqref{MOTeq} disregards two-body losses induced by collisions of MOT atoms with $D$-states ones, which can persist in the MOT region due to the trapping effect of the MOT quadrupole gradient. This approximation is justified by the large mismatch, up to two orders of magnitude, between the large volume occupied by the magnetically trapped $D$-state atoms and the one characteristic of the MOT, as we derive in the following. From Eq.\,\eqref{MOTeq} it is straightforward to verify that the steady-state  population of the MOT monotonically increases for decreasing $\gamma_{P\rightarrow {D}}$ values, up to the saturation value $\sqrt{\Gamma_{D}/\beta'_{MM}}$ for zero one-body losses, ideally allowed by the presence of red repumper lights. Correspondingly, the typical loading time also increases, owing to the reduced number of loss channels.

An accurate analysis of the MOT dynamics goes beyond the scope of the present work. Nonetheless, the results of Fig.\,\ref{CrLoading}(a) and (b) allow us to estimate the parameters $\gamma_{P\rightarrow {D}}$ and $\beta_{MM}$, under our working conditions. To this end, we proceed as follows.
We first fit the solution of the rate equation \eqref{MOTeq} to the datapoints of the MOT loading without red repumper light [Fig.\,\ref{CrLoading}(a)]. The best fit is obtained for vanishing $\beta'_{MM}$, meaning that we are not sensitive to 2-body collisions under these conditions.
By fixing $\beta_{MM}$, hence $\beta'_{MM}$, to the literature value $1.5\,10^{-9}$cm$^3\,$/s \cite{ChicireanuPhD}, we obtain the best-fit result $\Pi_P\gamma_{P\rightarrow {D}}=11.7(1)\,$Hz. The collisional volume is calculated from the \textit{in situ} size of the MOT derived from absorption images. 
From the knowledge of our MOT light absolute intensity within a $30$\% uncertainty, we obtain $\Pi_P=0.04(1)$, and hence $\gamma_{P\rightarrow {D}}=290(60)\,$Hz, which is in good agreement with previous findings \cite{ChicireanuPhD}.

We then move to the data taken in the presence of red repumpers [Fig.\,\ref{CrLoading}(b)], to which we fit the same model.
In this case, the one-body loss rate is determined by the decay to the $F=9/2$ manifold of the $^5D_4$ state, which is not addressed by our repumper lights. Based on the relative strength of the repumper transitions, consistently confirmed by our spectroscopy scans, we estimate this decay to be $1/4$ of the total one. By fixing the one-body loss rate to such a value, we obtain $\beta_{MM}=5\,10^{-10}\,$ cm$^3$/s. This is in reasonable agreement with, although slightly smaller than, the value reported in literature \cite{ChicireanuPhD} for analogous experimental settings.

The experimental routine followed to produce the data of Fig.\,\ref{CrLoading}(b) allows us to attain steady-state MOT containing up to $1.6 \,10^6$ atoms, a value upper limited by two-body light-assisted losses. 
A further sizable increase of the population of our cold $^{53}$Cr sample can be obtained by accumulating low-field seeking metastable $D$-state atoms within the  magnetic MOT quadrupole. 
Indeed, also for the fermionic Cr isotope, previous studies \cite{Chicireanu2006} suggest that $D$-state atoms feature two-body loss rates more than one order of magnitude smaller than the one induced by light-assisted collisions in the MOT.   
Similarly to what was already done in previous experiments \cite{Stuhler2001,Chicireanu2006}, we thus load the $^{53}$Cr MOT with no red repumpers [see Fig.\,\ref{CrLoading}(a)] for a variable time. We then apply a $30\,$ms-long pulse of the red repumpers onto the blue MOT, after which we monitor the atomic cloud through absorption imaging after a short time of flight. 
This allows us to reveal, besides the MOT cloud, also the population of magnetically trapped $D$-state atoms as a function of time. A typical loading curve, recorded under optimum MOT loading parameters [see Table \ref{Table2}], is presented in Fig.\,\ref{CrLoading}(c). 
Compared to the MOT one, the latter dynamics features completely different timescales and saturation values, see Fig.\,\ref{CrLoading}(a,b). In this case, by fitting the loading curve to a phenomenological exponential rise $N_{\infty} (1- e^{-t/\tau})$, we obtain a $1/e$ time $\tau=11.7(8)\,$s, and a steady-state number of $N_{\infty}=8.6(4)\,10^6$, corresponding to a loading rate of about $\Gamma=7.3(5) \, 10^5\,$atoms/s. Namely, in spite of a six-fold reduced loading rate compared to the MOT one, almost $10^7$ $D$-state $^{53}$Cr atoms can be collected within the magnetic trap, a number about $30$ times larger than the one of our steady-state MOT.

We now apply a more quantitative analysis on the results of Fig.\,\ref{CrLoading}(c), by exploiting the following model rate equation, coupled to Eq.\,\eqref{MOTeq}:
\begin{eqnarray}
N_{D}^{'}(t) = \eta \gamma_D N_{M}(t) -\tau_D^{-1} N_D(t) -\beta^{'}_{DD} N_{D}^2(t).
\label{Deq}
\end{eqnarray}
The first term accounts for the seeding of the quadrupole trap from the MOT, and the factor $\eta <1$ represents the fraction of $D$-state Zeeman sublevels that can be trapped by the quadrupole gradient. Magnetically trapped $D$-state atoms undergo one-body losses [second term of Eq.\,\eqref{Deq}] with a rate $\tau_D^{-1}$, and two-body losses [third term of Eq.\,\eqref{Deq}] with a rate $\beta'_{DD}=\beta_{DD}/V_{DD}$. $\beta_{DD}$ is the rate coefficient per unit volume, and $V_{DD}$ the collisional volume of the $D$-state cloud. The absence of the third red repumper during these measurements, hence the not complete repumping of metastable atoms, is taken into account by a scaling factor for the solution $N_{D}(t)$. Also in this case we neglect collisions between MOT and metastable Cr atoms. 

In order to quantify the inelastic parameters $\beta_{DD}$ and $\tau_D$ we proceed as follows.
We first derive the one-body loss rate $\tau_D^{-1}$ by monitoring the decay dynamics of magnetically trapped clouds for two different density conditions, realized under either an optimum or non-optimum loading of the magnetic trap. The latter is obtained by suppressing HF, TC and $R_3$ lights.
In both cases, we continuously load the magnetic trap for 4 s, after which we wait for a variable hold time with the MOT lights off. Then, by flashing the red repumpers for 30 ms, we recapture the atoms in the blue MOT, and we acquire an absorption image of the sample in time of flight.
An example of such a characterization is shown in Fig.\,\ref{Crdecay} (blue squares). 
A fit of these data to an exponentially decaying function yields $\tau_D=48(2)\,$s, irrespective of the initial $D$-state cloud density. 
We remark that such a lifetime is not affected by collisions with hot background Cr atoms, as $\tau_D$ does not vary with the presence or absence of the thermal atomic beam, which can be blocked by an in-vacuum shutter. 
The extracted $\tau_D$ is significantly larger than the $8\,$s lifetime measured in the Paris experiment \cite{ChicireanuPhD}, and it might be the key factor that allows us to obtain significantly larger $^{53}$Cr samples, containing almost ten times more $D$-state atoms.

By fixing $\tau_D$ and $\gamma_D$ to the previously determined values, a fit of the data in Fig.\,\ref{CrLoading}(c) to Eq.\,\eqref{Deq}, with only $\eta$ and $\beta'_{DD}$ as free parameters, yields  $\eta=0.25(1)$ and $\beta'_{DD}=5.6(2.2)\,10^{-9}\,$s$^{-1}$, respectively. 
The fitted value of $\eta$ corresponds to an average magnetic moment of $6.2(2)\,$MHz/G for the trapped $D$-state atoms. 
From this we can estimate the collisional volume $V_{DD}$, and thus obtain the two-body loss rate parameter per unit volume $\beta_{DD}$. 
Assuming the magnetically trapped atoms to be in thermal equilibrium at a temperature of 1/3 of the MOT one, as shown by previous work on $^{52}$Cr \cite{Stuhler2001}, we derive $\beta_{DD}=5.7(2.3)\,10^{-11}\,$cm$^3$/s, a value close to the one we measured for the bosonic $^{52}$Cr isotope, of $7\,10^{-11}\,$cm$^3$/s \cite{Neri2019}. 
We emphasize that the extracted value of $\beta'_{DD}$ is almost three orders of magnitude lower than the light-assisted loss term $\beta'_{MM}$ obtained for the MOT dynamics. This allows us to greatly increase the number of cold $^{53}$Cr atoms, by letting them fall into the $D$ states and accumulate for a few seconds in the magnetic quadrupole, and by recapturing them in the MOT only at the end upon shining the red repumpers, while setting the cooling light at an optimum detuning of $-5.0\,\Gamma_{Cr}$.

\begin{figure}[t!]
\begin{center}
\includegraphics[width= 8cm]{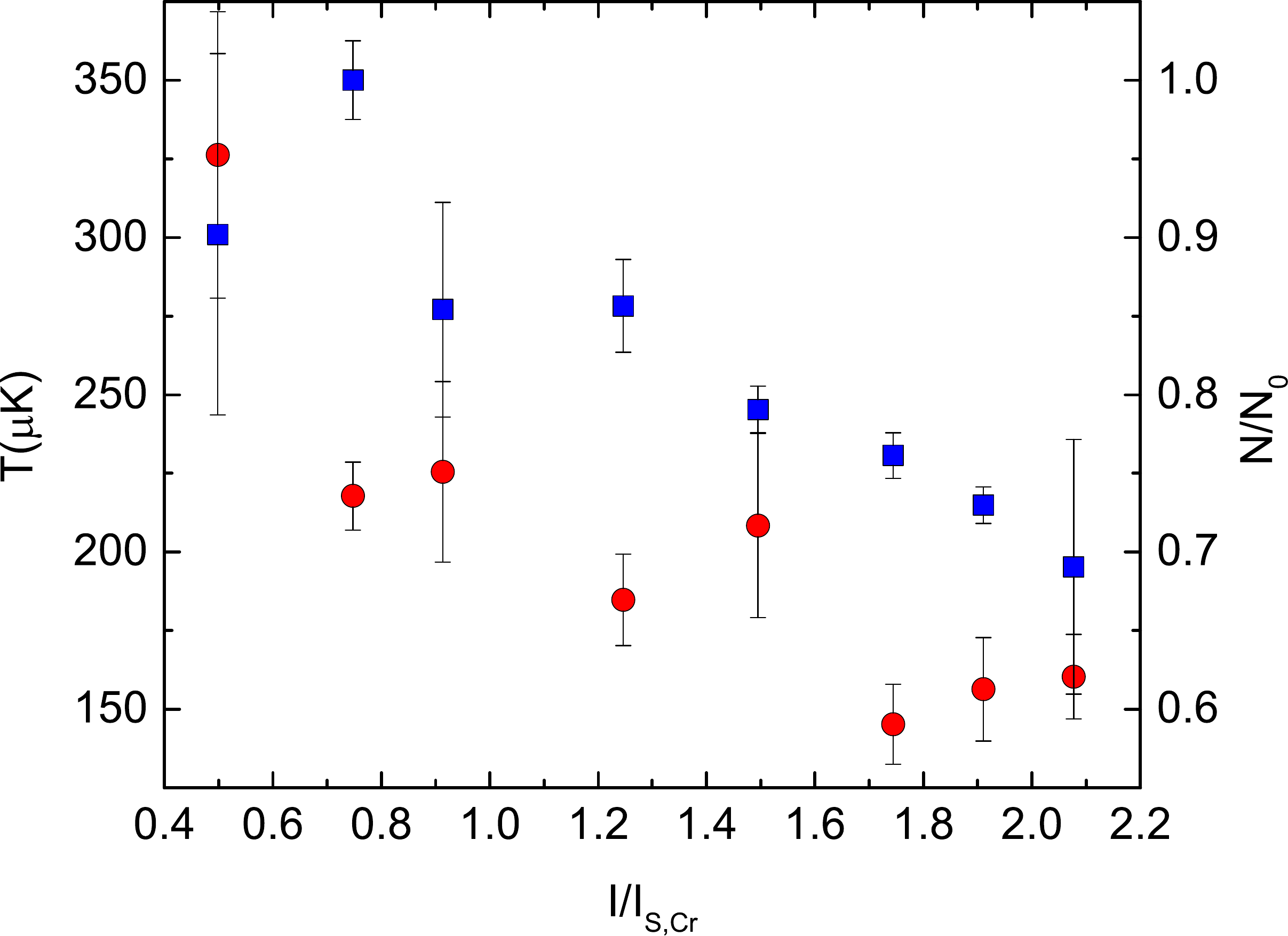}
\caption{Temperature of the Cr cloud (red circles, left axis) and surviving atomic fraction (blue squares, right axis) after a $10\,$ms-long CMOT stage, as a function of the single-beam cooling intensity. For each probed intensity, the cloud temperature is derived by the cloud size imaged at two different times of flight. Here, the cooling detuning was set to $\delta= -1.6\, \Gamma_{Cr}$, all other beam parameters were fixed to the values in Table \ref{Table2}.}
\label{CMOTCr}
\end{center}
\end{figure}

This procedure enables to maximize the atom number, but it results in relatively high MOT cloud temperatures of $T=530(70)\,\mu$K, about four times higher than the Doppler one, $T_D=124\,\mu$K.
In order to decrease the temperature of our sample, we apply a final $10\,$ms CMOT stage, reducing the cooling light detuning to $\delta=-1.4 \,\Gamma_{Cr}$. 
By varying the cooling intensity applied at this stage, while keeping constant all the other parameters, we have pinpointed the best conditions that minimize the cloud temperature while not causing substantial atom losses, see Fig.\,\ref{CMOTCr}.
In particular, for a single-beam MOT intensity of about $1.8\,I_{S, Cr}$, we can reach temperatures of $145(5)\,\mu$K, less than 20$\%$ above the Doppler limit, with a surviving fraction exceeding 75$\%$. 
The CMOT stage sizeably shrinks the atomic cloud, decreasing its \textit{in situ} $1/e$ radius from $270(30)\,\mu$m down to $170(15)\,\mu$m. Correspondingly, it greatly increases the cloud peak density, reaching values of about $10^{11}\,$atoms/cm$^3$. 
Our final $^{53}$Cr CMOT temperature is comparable to previously reported values \cite{Chicireanu2006}, whereas the peak densities featured by our clouds are considerably larger than those achieved by the Paris group.
   
\section{EXPERIMENTAL RESULTS: DOUBLE SPECIES} \label{doublespecies}

Having detailed our strategies to produce individual Li and Cr cold clouds, we now discuss how the two species, collected within the same MOT region, affect one another. 
First of all, we have checked that neither the Li $671\,$nm light nor the Li atomic beam affects the Cr sample. Similarly, we have found that application of the Cr $425\,$nm lights, as well as of the red repumpers, do not modify the features of our Li MOT. Moreover, while the simultaneous presence of the two ZS fields modifies the quadrupole center and the position of each MOT, the optimum loading performances can be easily retrieved by slightly realigning the optical beams.  

We then move to inspect the mutual effect of Li and Cr atoms in the cold regime, when simultaneously collected within a double-species MOT. In particular, since the number of Li atoms is typically two orders of magnitude higher than the Cr one, we focus on the loading dynamics of the chromium sample. Although, in practice, the two clouds can be easily displaced one from the other by tweaking the polarization and alignment of the  beams, we intentionally investigate the Cr cloud, loaded into the MOT and within the magnetic trap, under the worst conditions possible, namely once its overlap with a Li MOT, of roughly $10^8$ atoms at mK temperatures, is maximized.
We have first monitored the Cr MOT loading, both without and with red repumpers, in the presence of an underlying Li cloud. No appreciable reduction of our Cr MOT loading efficiencies is observed in either case. This fact represents a major advantage of our novel biatomic system: in contrast with the case of other available ones [see, e.g., Ref.\,\cite{Vaidya2015} for the Rb-Yb case], we can conveniently perform a simultaneous loading of the two species within the same spatial region. This  automatically provides an optimum starting point for the subsequent optical trapping and evaporation stages, and it considerably shortens the experimental cycle.
\begin{figure}[t!]
\begin{center}
\includegraphics[width= 8cm]{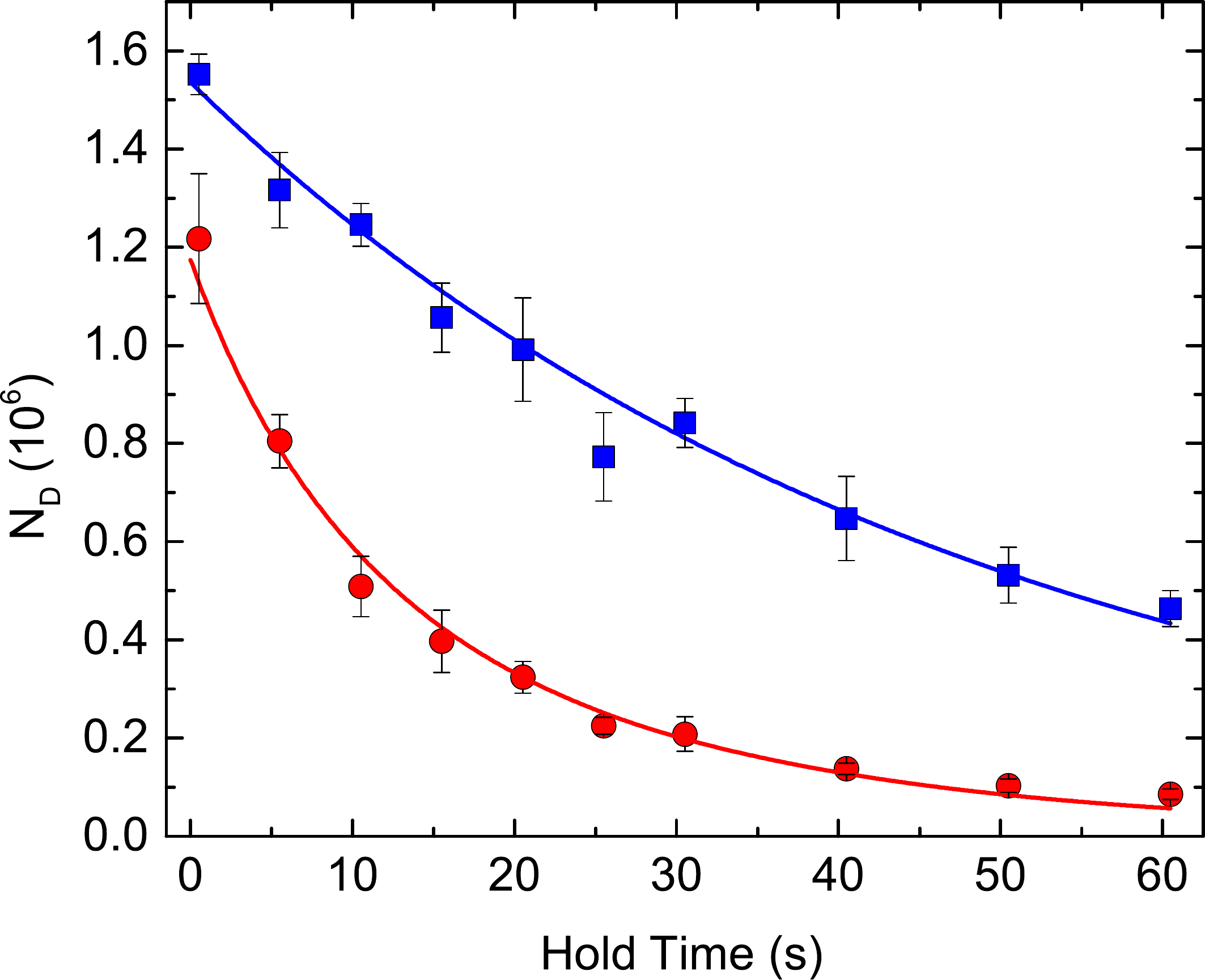}
\caption{Stability of a magnetically trapped Cr $D$-state cloud accumulated for 3 s within the MOT quadrupole, with (red circles) and without (blue squares) a $^6$Li MOT of about $10^8$ atoms and a peak density of $2\,10^9\,$atoms/cm$^3$ superimposed to the quadrupole center. The single Cr cloud features a dynamics that is excellently fitted by a single exponential decay, with $1/e$ time of $\tau_D= 48(2)\,$s (blue solid curve). The presence of the Li cloud reduces, although not dramatically the Cr $1/e$ lifetime to $13.0(7)\,$s (red solid curve).}
\label{Crdecay}
\end{center}
\end{figure}

Finally, we discuss how the presence of the Li cloud affects the stability of magnetically trapped samples of $D$-state Cr atoms. Also for this measurement, we intentionally put ourselves in the least favourable conditions, by first loading the Li MOT centered on the magnetic quadrupole, and then by collecting a relatively small cloud, of less than two million metastable $^{53}$Cr atoms, within the subsequent three seconds. Once the Cr cloud has been produced in the magnetic trap, we turn off the Cr lights, wait for a variable time, after which we turn on both red repumpers and blue MOT beams for  $30\,$ms, and finally apply a 3-ms-long CMOT stage on chromium. 
We then let both the Li and Cr clouds ballistically expand for $0.6\,$ms, after which we acquire two consecutive absorption images. 
The resulting decay curve for Cr is shown in Fig.\,\ref{Crdecay}, red circles, and it is contrasted with the dynamics of Cr atoms in the absence of the Li MOT, blue squares, obtained through the same routine while keeping the Li oven shutter closed. 

By comparing the two data sets, one can see how the Li MOT causes a moderate but detectable decrease, of about 25$\%$, of the quadrupole population loaded after the  $3\,$s loading. Parallel to this, the $1/e$ lifetime of the magnetically trapped cloud drops, although not dramatically, from $48(2)$ in the single species case, down to $13.0(7)\,$s when the Li MOT is present. 
We remark that two-body losses connected with intra-species collisions between $D$-state atoms are negligible for these Cr densities: identical decay times, in absence of Li atoms, are indeed observed for a Cr cloud with a ten-fold lower atom number [not shown in the figure]. 
The Li sample features a decay with a typical lifetime of about $25\,$s, unaffected by the presence of the Cr cloud. Although we cannot exclude other loss mechanisms, we ascribe the observed reduction of the Cr lifetime to multiple spin-exchange collisions between $D$-state Cr with Li atoms. These processes, which release energies on the order of a few hundreds of $\mu$K, do not cause Li losses from the MOT, but promote the Cr atoms towards high-field seeking Zeeman sublevels of the $^5D_4$ and $^5D_3$ states, causing their loss from the magnetic trap. 
By applying the same model as the one exploited for Cr alone, see again Eq.\,\eqref{Deq}, to the Cr decay data in Fig.\,\ref{Crdecay}, we extract a value of $\beta'_{Cr,Li}=9.8(4)\,10^{-10}\,$Hz for the inter-species two-body loss rate. 
With a collisional volume estimated from the in situ images of the lithium cloud and from the volume of $D$-state Cr atoms derived in Section \ref{singlespecies} , this result yields a value of $\beta_{Cr,Li}=5.0(3)\,10^{-11}\,$cm$^3$/s for the rate coefficient per unit volume.

Our study indicates that the protocols devised for optimum Cr loading, relying on the accumulation of $D$-state atoms into the MOT quadrupole, can be suitably employed also to produce cold Cr-Li mixtures. 
In particular, we have verified that a sequential loading of the two species [both with the Li MOT produced first, and vice versa] leads to samples analogous to those obtained with the simplest simultaneous loading scheme, which enables to create clouds of about $1.5\,10^8$ Li and $4 \, 10^6$ Cr atoms on a $5\,$s-long experimental cycle.  

\section{CONCLUSIONS AND OUTLOOK}\label{outlook}
In conclusion, we have produced a novel mixture of $^6$Li and $^{53}$Cr fermionic atoms in the cold regime. For lithium, our setup enables to realize clouds of more than 10$^8$ atoms at sub-Doppler temperatures within a few seconds cycle time. 
Most importantly, we have demonstrated the possibility to realize comparatively large clouds of several millions of $^{53}$Cr near the Doppler temperature, by relying on the accumulation of dilute samples of metastable $D$-state atoms within the MOT magnetic quadrupole, followed by a final CMOT stage. 
The presence of a large cloud of about 10$^8$ Li atoms, overlapped with the Cr one, does not dramatically compromise the loading performance observed for the single Cr gas, allowing for a simultaneous loading of the two species with a fast cycle time.
 
This appears as a very promising starting point to implement evaporative cooling of $^6$Li and sympathetic cooling of $^{53}$Cr atoms within an optical dipole trap, following a  strategy conceptually analogous to the one already successfully implemented for the $^6$Li-$^{40}$K mixture \cite{Spiegelhalder2010}. 
In particular, we expect the small size of the Cr MOT to enable a very large collection efficiency \cite{Naylor2015} within our recently realized optical trapping potential \cite{Simonelli2019}. 
Although the scattering properties of the Cr-Li system are totally unknown, yet, as mentioned already in Section \ref{intro}, the complex hyperfine and Zeeman structure of chromium, combined with its highly magnetic character, is expected to guarantee a relatively rich spectrum of heteronuclear Feshbach resonances \cite{Simonipriv}. These could be employed to enhance the interspecies scattering cross section, and thus to optimize the sympathetic cooling efficiency even when small background Cr-Li scattering lengths were found.   
Finally, we also emphasize that the realization of large and cold $^{53}$Cr clouds, demonstrated in this work, may also enable, in the future, to evaporatively cool down to quantum degeneracy binary spin mixtures of fermionic chromium, with no need to rely on inter-species collisions with a second atomic component.

\begin{acknowledgments}
We thank T. Pfau, B. Laburthe-Tolra, L. Vernac and E. Maréchal for insightful discussions and for sharing their know-how on chromium. 
We acknowledge A. Simoni and D. Petrov for stimulating discussions and for sharing their data on the two- and three-body properties Cr-Li mixtures. 
Special thanks to F. DiNoia, A. Cosco, M. Seminara and M. Jag for their contribution in the early stage of the experiment, and to the whole Quantum Gases Group at LENS, in particular G. Roati, for the continuous support.
This work was supported by the ERC through grant no.\:637738 PoLiChroM and by the Italian MIUR through the FARE grant no.\:R168HMHFYM.
\end{acknowledgments}


\vskip50pt

%

\begin{thebibliography}{75}
\bibitem [{CrL()}]{CrLi_sources_note}%
  \BibitemOpen
  \href@noop {} {}\bibinfo {note} {The lithium source contains 10\,g, 95\%
  enriched \Li{} provided by AMES lab, packaged in glass ampoules with inert
  gas. The chromium source contains ca. 10\,g of natural chromium (99.99\%
  clean, abundance 9.5\% $^{53}$Cr) provided by {G}oodfellow in lumps of
  $0.7-3.5$\,mm. The liners and end caps are produced by Keratec Advanced
  Materials SA. The Zirconia based glue for the end cap is Duralco Resbond 904
  from Cotronics Corp.}\BibitemShut {Stop}%
\end{thebibliography}

%

\end{document}